\documentclass{article}
\usepackage{amscd,amsmath,amssymb,latexsym,bbm}
\def\DH{\rm I\kern-1.5pt\rm H\kern-1.5pt\rm I}

\newcommand{\sfrac}[2]{{\textstyle\frac{#1}{#2}}}

\def\DR{\rm I\kern-1.45pt\rm R}
\def\DC{\kern2pt {\hbox{\sqi I}}\kern-4.2pt\rm C}

\newcommand{\ba}{\begin{array}}
\newcommand{\ea}{\end{array}}
\newcommand{\be}{\begin{equation}}
\newcommand{\ee}{\end{equation}}
\newcommand{\bea}{\begin{eqnarray}}
\newcommand{\eea}{\end{eqnarray}}
\newcommand{\bi}{\begin{itemize}}
\newcommand{\ei}{\end{itemize}}

\newcommand{\bs}{\mbox{\boldmath $\sigma$}}

\newcommand\ointint{\begingroup
\displaystyle \unitlength 1pt
\int\mkern-12.2mu
\begin{picture}(0,3)
\put(0,1){\oval(5,5)}
\end{picture}\mkern 6mu
\endgroup}

\newcommand{\R}{\mathbb R}
\newcommand{\Z}{\mathbb Z}
\newcommand{\NN}{\mathbb N}

\begin{document}
\thispagestyle{empty}
\begin{flushright}
ITP--UH--07/10\\
\end{flushright}
\medskip

\begin{center}
{\bf \Large Action-angle variables for dihedral systems on the circle}\\
\vspace{0.5 cm} {\large
Olaf Lechtenfeld$\;^{a}$,
Armen Nersessian$\;^{b}$ and
Vahagn Yeghikyan$\;^{b}$}
\end{center}
\begin{center}
$\;^a${\sl Leibniz Universit\"at Hannover,
Appelstr. 2, D-30167 Hannover, Germany}\\
$\;^b${\sl Yerevan State University,
1 Alex Manoogian St., Yerevan, 0025, Armenia}
\end{center}

\begin{abstract}
A nonrelativistic particle on a circle and subject to a $\cos^{-2}(k\varphi)$
potential is related to the two-dimensional (dihedral) Coxeter system $I_2(k)$,
for $k\in\NN$. For such `dihedral systems' we construct the action-angle
variables and establish a local equivalence with a free particle on the circle.
We perform the quantization of these systems in the action-angle variables
and discuss the supersymmetric extension of this procedure. By allowing radial
motion one obtains related two-dimensional systems, including $A_2$, $BC_2$ and
$G_2$ three-particle rational Calogero models on~$\R$, which we also analyze.
\end{abstract}

\section{Introduction}
One of the most important theorems in the theory of integrable
systems is the Liouville theorem~\cite{arnold}. It investigates
$n$-dimensional systems which have $n$ mutually commuting constants
of motion $F_1{\equiv}H,\ldots, F_n$: $\{F_i, F_j\}=0$ with $i,j=1,\ldots n$.
Besides establishing of the classical integrability of such a system,
the theorem states that if the level surface $M_f=\left( (p_i,q_i):
F_i=\textrm{const}\right)$ is a {\sl compact and connected manifold\/},
then it is diffeomorphic to the $n$-dimensional torus $T^n$. The natural
coordinates $\Phi_i$ parameterizing this torus obey free equations of motion.
Together with their conjugate momenta $I_i$, they fully parameterize the phase
space and are called ``action-angle'' variables. These momenta are conserved
and only depend on the constants of motion, i.e.~$I_i=I_i(F)$. Hence, we may
perform a canonical transformation $(p_i,q_j)\mapsto(I_i,\Phi_j)$ to the new
variables, in which the Hamiltonian $H$ depends only on the
(constant) action variables~$I_i$. Consequently, the equation of motion reads

\be
\frac{dI_i}{dt}=0,\quad \frac{d\Phi_i}{dt}=\frac{\partial
{H}(I)}{\partial I_i}=: \omega_i(I),\qquad \{ I_i, \Phi_j
\}=\delta_{ij},\qquad  \Phi_i\in [0,2\pi ),\qquad I_i\in\R_+.
 \ee
The formulation of an integrable system in these variables gives us a
comprehensive geometric description of its dynamics and is a useful
tool for developing perturbation theory~\cite{arnold,goldstein}.
For these reasons, action-angle variables have been widely exploited in
celestial mechanics since the 19th~century and play a central role in
the Bohr-Sommerfeld semiclassical quantization.
Furthermore, their use gives us a criterion for the (non)equivalence of
two integrable systems, since the latter are characterized by two data:

-- the functional dependence $H(I_i)$,

-- the domain of the action variables, $I_i\in[\beta^{-}_i,\beta^{+}_i]$.\\
During the last half-century, numerous new finite-dimensional integrable
systems with a rich mathematical structure have been discovered. Some examples
are the (pseudo)spherical generalizations of the Kepler system~\cite{sch} and
of the oscillator~\cite{higgs} as well as Calogero-type multi-particle systems
\cite{calogero}, which may be coupled to monopoles or instantons.
Yet, to our knowledge, explicit expressions for their action-angle variables
are absent in the literature, even for those systems which admit a
separation of (spatial) variables.

In this Note we construct the action-angle variables for the dihedral systems
on a circle, which are defined by the Hamiltonian
\be
H(p,q)={\cal I}(p_\varphi,\varphi|k)\ =\ 
\sfrac{1}{2} p_\varphi^2\ +\ V_k(\varphi) ,
\label{ham}\ee
with canonical variables $\{p_\varphi,\varphi\}=1$ and the potential
\be
V_k(\varphi)\ =\ \sum_{\ell=0}^{k-1}\frac{1}{({\bf a}_\ell\cdot {\bf n})^2}
\qquad\textrm{where}\quad
{\bf n}=\begin{pmatrix} \cos\varphi \\ \sin\varphi \end{pmatrix}\label{v0}
\ee
and ${\bf a}_\ell$ are the positive roots of a two-dimensional Coxeter system
$I_2(k)$ called dihedral system. The full set of roots forms a regular star
shape with an angular separation of $\pi/k$.
Since the dihedral symmetry (discrete planar rotations and reflections)
relates the root lengths
as $|{\bf a}_\ell|^2=|{\bf a}_{\ell+2}|^2$, for odd~$k$ all roots have the same
length, say $\alpha_0$, while for even~$k$ we may put
$|{\bf a}_{\textrm{even}}|=\alpha_1$ and $|{\bf a}_{\textrm{odd}}|=\alpha_2$.
Clearly, we have to distinguish between $k$ being even or odd.
As ${\bf a}_\ell{\cdot}{\bf n}$ is proportional
to $\cos(\varphi{-}\frac{\ell\pi}{k})$,
the finite sums can be performed by consulting a handbook~\cite{PBM} to obtain
\bea
V_k(\varphi) &=&
\frac{k^2\alpha_0^2}{ 2\cos^2 k\varphi}
\qquad\qquad\qquad\qquad\ {\rm for}\; k=2k'{+}1,
\label{hggodd}\\[3mm]
V_k(\varphi) &=&
\frac{(k'\alpha_1)^2}{2\cos^2 k'\varphi}\ +\
\frac{(k'\alpha_2)^2}{2\sin^2 k'\varphi}
\qquad {\rm for }\; k=2k',
\label{hggeven}\eea
with $k'\in\NN$. Hence, the odd systems feature one coupling ($\alpha_0$),
while the even ones allow for two ($\alpha_1$, $\alpha_2$), all naturally
positive. For $\alpha_1{=}\alpha_2$, the even potential attains the same form
as the odd one. 
Due to the singularities at $\varphi=\sfrac{\pi}{2}$ mod~$\pi$ for odd~$k$ 
and at $\varphi=0$ mod~$\pi$ for even~$k$, the configuration space decomposes
into $2k$~disjoint pieces, which are equivalent since related via translation
by $\Delta\varphi=\frac{\pi}{k}$. 
We shall restrict ourselves to just one of them.

Formulating these systems in terms of action-angle-variables, we shall find
that both types are locally equivalent to the free particle on the circle.
Besides, we establish a global equivalence between systems~(\ref{hggodd})
and~(\ref{hggeven}) for $\alpha_0=\alpha_1+\alpha_2$ and
$2k_{\textrm{odd}}=k_{\textrm{even}}$.
We shall demonstrate that, after restricting to one of the $2k$ branches,
these systems are equivalently quantized in their
action-angle variables or initial coordinates.
We shall also present a supersymmetrization of the action-angle variable scheme
for the dihedral systems and its relation to the supergeneralization of
the Liouville theorem.

Finally, we shall shall enlarge the configuration space to~$\R^2$ by
adding a radial degree of freedom to the circular motion. The ensueing
two-dimensional systems represent three-particle rational Calogero models
after separation of their center-of-mass motion. For small values of~$k$,
the Coxeter roots belong to a rank-two Lie algebra~$\cal G$, which labels
the corresponding Calogero model~\cite{algebras}:
\be
\begin{tabular}{|c|cccc|}
\hline
$k$ & 2 & 3 & 4 & 6 \\
\hline
$\cal G$ & $\ D_2=A_1{\oplus}A_1\ $ & $A_2$ & $\qquad BC_2\qquad$ & $G_2$ \\
\hline
\end{tabular}
\label{tab}\ee
In particular, this shall allow us to prove the global equivalence of
the $A_2$ and $G_2$ rational Calogero models and their local equivalence
to a free particle in the plane.

\section{Action-angle variables}
In this Section we construct the action-angle variables for the
systems defined by the potentials (\ref{hggodd}) and
(\ref{hggeven}). We follow the general prescription given in
\cite{arnold}. Being one-dimensional, our systems feature the
Hamiltonian as their single constant of motion. To construct the
action variable, we should fix the level surface of the Hamiltonian,
$H(p_\varphi,\varphi)={\cal I}(I)=h$, and introduce the generating
function $S(h,\varphi)$ for the canonical transformation
$(p_\varphi,\varphi)\mapsto({ I},\Phi)$ via \be
S(h,\varphi)=\int_{\varphi_0}^\varphi p_\varphi(h,\varphi')\
d\varphi' =\int_{\varphi_0}^\varphi \sqrt{2(h-V_k(\varphi'))}\
d\varphi' . \label{sdef}\ee
The full period integral yields the
action variable,
 \be I(h)=\frac{1}{2\pi}\ointint\
p_\varphi(h,\varphi')\ d\varphi' =\frac{1}{2\pi}\ointint\
\sqrt{2(h-V_k(\varphi'))}\ d\varphi' , \label{actionvar}\ee
while
the angular variable $\Phi$ arises from \be \Phi(h,\varphi)=
\frac{\partial S}{\partial I}= \frac{dh}{dI}\frac{\partial S}{\partial h} =2\pi
\int_{\varphi_0}^\varphi \frac{d\varphi'}{\sqrt{2(h-V_k(\varphi'))}}
\biggm/ \ointint\ \frac{d\varphi'}{\sqrt{2(h-V_k(\varphi'))}}.
\label{anglevar}\ee The parity of parameter $k$ does not play any
role in our derivation.\footnote{ Formally, $k$ need not even be an
integer. In such a case, however, the system lives on the infinite
cover~$\R$ of the circle.} Surely, in the limit $\alpha_2\to0$ the
system (\ref{hggeven}) looks like system~(\ref{hggodd}). In our
construction, however, it is essential to keep both $\alpha_{1}$ and
$\alpha_{2}$ non-vanishing. For this reason we shall derive the
action-angle variables of the even and odd systems separately.

\subsubsection*{Systems with odd $k$}
We concentrate on the range $\varphi\in[-\sfrac{\pi}{2k},\sfrac{\pi}{2k}]$
for the configuration variable.
Inserting (\ref{hggodd}) into (\ref{actionvar}), we obtain
\be
I=\frac{\sqrt{2}}{\pi}\int_{\varphi_-}^{\varphi_+}d\varphi'
\sqrt{h-\frac{k^2\alpha_0^2}{2\cos^2{k\varphi'}}},
\label{idef}\ee
where the reflection points $\varphi_\pm(h)$ follow from
\be
2h\,\cos^2{k\varphi}_{\pm} = k^2\alpha_0^2 .
\ee
Calculating the definite integral (\ref{idef}), we find
\be
I=\frac{1}{k}\sqrt{2h}-\alpha_0
\qquad \Rightarrow\qquad
{\cal I}=\frac{k^2}{2}(I+\alpha_0)^2
\label{ch}\ee
and thus get
\be
\frac{d h}{d I}=k^2(I+\alpha_0)=k\sqrt{2h} .
\ee
At the potential mimimum ($\varphi{=}0$), we have $I=0$ but
$h=h_{\textrm{min}}=\sfrac12k^2\alpha_0^2$.
To compute the angular variable $\Phi$ we employ (\ref{anglevar})
with $\varphi_0{=}0$ and get
\be
\Phi= \frac{dh}{dI} \frac{1}{k\sqrt{2h}}
\int_0^{x(\varphi)}\!\!\frac{dx'}{\sqrt{1-{x'}^2}}=\arcsin{x(\varphi)}
\qquad\textrm{where}\quad
x(\varphi):=\frac{\sqrt{2h}}{\sqrt{2h-k^2\alpha_0^2}}\sin{k\varphi}.
\ee
Hence, the canonical transformation to the action-angle
variables looks as follows,
\be
\left(p_\varphi, \varphi\right)\mapsto
\Bigl(I=\sfrac{1}{k}\sqrt{2{\cal I}(p_\varphi,\varphi)}-\alpha_0\ ,\
\Phi=\arcsin\Bigl\{\sfrac{\sqrt{2{\cal I}(p_\varphi,\varphi)}}
{\sqrt{2{\cal I}(p_\varphi,\varphi)-k^2\alpha_0^2}}\sin{k\varphi}\Bigr\}\Bigr),
\ee
where ${\cal I}(p_\varphi,\varphi)$ is given by (\ref{ham}) and~(\ref{hggodd}).
When the particle makes one cycle (the variable $x$ runs from $-1$ to $1$
and back), the variable $\Phi$ advances by $2\pi$ as expected.
In these variables the Hamiltonian is given by the second expression
in~(\ref{ch}).
For completeness, the inverse transformation 
$(I,\Phi)\mapsto(p_\phi,\phi)$ reads
\be
\varphi=\frac{1}{k}\arcsin
\Bigl\{{\sfrac{\sqrt{I^2+2I \alpha_0 }}{{I}+\alpha_0}\sin{{\Phi}}}\Bigr\},
\qquad p_\varphi=k{(I{+}\alpha_0)}
\sqrt{\sfrac{{I}^2+2I\alpha_0}{(I+\alpha_0)^2+(\alpha_0\tan{{\Phi}})^2}}.
\ee

Performing the trivial canonical transformation
$(I,\Phi)\mapsto({\tilde I} =I{+}\alpha_0, \Phi)$,
we get
\be
{\cal I}=\frac{k^2}{2}{\tilde I}^2 \quad\textrm{with}\quad
\{{\tilde I},{\Phi} \}=1,\qquad\textrm{where}\quad
{\Phi}\in [0,2\pi)\quad\textrm{and}\quad
{\tilde I}\in[\alpha_0,\infty).
\label{shift}\ee
This system can be interpreted as a free particle particle of mass $k^2$
moving on a circle with unit radius.
Equivalently, it describes a free particle of unit mass moving on a circle
with radius $1/k$. However, we can speak about {\it local\/}
equivalence only, since the above redefinition changes the domain of
the action variable from $[0,\infty)$ to $[\alpha_0,\infty)$!

\subsubsection*{Systems with even $k$}

For the case (\ref{hggeven}), i.e. $k=2k'$,
we select $\varphi\in[0,\frac{\pi}{k}]$.
The action variable is now slightly harder to compute,
\be
I(h)= \frac{\sqrt{2}}{\pi}\int^{\varphi_{+}}_{\varphi_{-}}d\varphi'
\sqrt{h-\frac{{k'}^2\alpha_1^2}{2\cos^2{k'\varphi'}}
-\frac{{k'}^2\alpha_2^2}{2\sin^2{k'\varphi'}}}
= \frac{\sqrt{2h}a^2}{k\pi}\int\limits_{-1}^1
\frac{\sqrt{1-x^2}\ dx}{1-\left(ax+b\right)^2},
\label{idef1} \ee
where
\be
a=\sqrt{1-\frac{{k'}^2(\alpha_1^2+\alpha_2^2)}{h}
+\frac{{k'}^4(\alpha _1^2-\alpha _2^2)^2}{4h^2}},
\qquad
b=\frac{{k'}^2(\alpha_2^2-\alpha_1^2)}{2h},
\qquad
x=\frac{1}{a}\bigl[\cos{2k'\varphi'}-b\bigr],
\label{definitions} \ee
and the turning points $\varphi_\pm(h)$ derive from
\be
2h\,\sin{k'\varphi}_\pm\cos{k'\varphi}_\pm
={k'}^2(\alpha_1^2\tan{k'\varphi}_\pm+\alpha_2^2\cot{k'\varphi}_\pm)
\qquad\textrm{with}\quad \alpha_1,\alpha_2>0.
\ee
The last integral in (\ref{idef1}) can be calculated by
standard methods (see Appendix) \cite{fiht}:
\be
\int\limits_{-1}^1\frac{\sqrt{1-x^2}\ dx}{1-\left(ax+b\right)^2}=
\frac{\pi}{2a^2}\Bigl(2-\sqrt{(b-1)^2-a^2}-\sqrt{(b+1)^2-a^2}\Bigr) ,
\label{idef2}\ee
thus,
\be
I=\frac{1}{k'}\sqrt{2h}-(\alpha _1+\alpha _2)
\qquad\Rightarrow\qquad
{\cal I}=\frac{{k'}^2}{2} \bigl(I+(\alpha_1+\alpha_2)\bigr)^2.
\label{che}\ee
Similarly, the angular variable becomes
\be
\Phi=\sfrac12\arcsin\Bigl\{\sfrac{1}{a}\bigl[\cos{2k'\varphi}
+b\bigr]\Bigr\},
\label{fig} \ee
where  $a$ and $b$ are defined by the expressions (\ref{definitions}), 
and $h$ should be replaced by ${\cal I}(p_\varphi,\varphi)$.
In these variables the Hamiltonian is displayed by the second expression
in (\ref{che}).
The inverse transformations $(I,\Phi)\mapsto(p_\varphi,\varphi)$ looks as
follows,
\be
\phi=\frac{1}{2k'}\arccos\Bigl\{a\sin{2{\Phi}}
-b\Bigr\},\qquad
p_\varphi=k'\sqrt{({I}+\alpha_1+\alpha_2)^2
-\sfrac{2\alpha_1^2}{1+b+a\sin{2\Phi}}-\sfrac{2\alpha_2^2}{1+b-a\sin{2\Phi}}},
\ee
where the quantities $a$ and $b$ take the form
\be
a=\sqrt{\Bigl[1-\bigl(
\sfrac{\alpha_1+\alpha_2}{{I}+\alpha_1+\alpha_2}\bigr)^2\Bigr]
\Bigl[1-\bigl(\sfrac{\alpha_1-\alpha_2}{{I}+\alpha_1+\alpha_2}\bigr)^2\Bigr]}, 
\qquad
b=\frac{k'^2(\alpha_2^2{-}\alpha_1^2)}{({I}{+}\alpha_1{+}\alpha_2)^2}.
\ee

Similar to the odd case, we perform a trivial
canonical transformation
$(I,\Phi)\mapsto({\tilde I}=I{+}\alpha_1{+}\alpha_2,\Phi)$
and arrive at
\be
{\cal I}=\frac{{k'}^2}{2}{\tilde I}^2 \quad\textrm{with}\quad
\{{\tilde I},{\Phi}\}=1,\qquad\textrm{where}\quad
{\Phi}\in [0,2\pi) \quad\textrm{and}\quad
{\tilde I}\in[\alpha_1{+}\alpha_2,\infty).
\label{shift1}\ee
So, also the even system (\ref{hggeven}) is {\it locally\/}
equivalent to a free particle of mass ${k'}^2$ moving
on the circle with unit radius (or a free particle of
unit mass moving on the circle with radius $1/{k'}$).
Like in the odd case, the equivalence is not global,
since the above shift of the action variable changes its range
from $[0,\infty)$ to $[\alpha_1{+}\alpha_2,\infty)$.

Comparing the results (\ref{shift}) and (\ref{shift1}),
obtained by a canonical transformation from (\ref{hggodd}) and
(\ref{hggeven}), respectively, we conclude that they differ in
the ``mass'' of the (locally equivalent) free particle as well as
in the domain of the momentum (action) variable. Thus, in general,
all systems can be distinguished globally. Interestingly, however,
any odd system $(k_{\textrm{odd}};\alpha_0)$ matches {\it globally\/}
to a one-parameter family system of even systems
$(k_{\textrm{even}};\alpha_1,\alpha_2)$ by the equivalence
\be
(k_{\textrm{odd}};\alpha_0)\ \sim\
(2k_{\textrm{odd}};\beta,\alpha_0{-}\beta)
\qquad\textrm{with}\quad 0<\beta<\alpha_0.
\ee

\subsubsection*{Quantization}
In the action-angle variables $(I,\Phi)$ derived above it
is quite simple to quantize the dihedral systems (\ref{hggodd}) and
(\ref{hggeven}) \`a la Bohr-Sommerfeld:
\be
I\mapsto{\widehat I}=\frac{\hbar}{\imath}\frac{\partial}{\partial \Phi},\qquad
\Psi_n=\sfrac{1}{\sqrt{2\pi}}{\rm e}^{\imath n\Phi}
\quad\textrm{for}\quad n\in\Z \qquad\Rightarrow\qquad
{\widehat I}\,\Psi_n(\Phi)=n\hbar\,\Psi_n.
\ee
The energy spectra of the Hamiltonians (\ref{ch}) and~(\ref{che}) then read,
respectively,
\be
E_n(k\ \textrm{odd}) =\sfrac12 k^2 (n\hbar+\alpha_0)^2 \qquad\textrm{and}\qquad
E_n(k\ \textrm{even})=\sfrac18 k^2 (n\hbar+\alpha_1{+}\alpha_2)^2 .
\label{spectra} \ee
This agrees with the literature~\cite{flugge,barut,ttw,quesne}, 
where the Schr\"odinger equation for our potential (\ref{hggodd}) or~(\ref{hggeven})
is known as the (first) P\"oschl-Teller equation, whose (normalizable) solutions 
are given in terms of trigonometric and hypergeometric functions.
Quantization in $(I,\Phi)$ variables gives rise to subtleties, however, due to
singular behavior at $I{=}0$~\cite{kastrup}.

\section{Supersymmetric extension}
The supergeometric generalization of the Liouville theorem has been known for
many years~\cite{shander}. For our context of one-dimensional supersymmetric
mechanics, we follow here the construction of action-angle (super)variables as
presented in~\cite{khud}.
Let us have ${\cal N}{=}2M$ one-dimensional supersymmetric mechanics defined
on a $(2|2M)$-dimensional phase superspace, coordinatized by
$(p_\varphi,\varphi|\theta^\alpha,\overline{\theta}^\beta)$.
The supersymmetry algebra reads
\be
\{Q^\alpha,\overline{Q}^\beta\}=2\delta^{\alpha\beta}H_s
\qquad\textrm{and}\qquad
\{Q^\alpha,H_s\}=\{\overline{Q}^\beta,H_s\}=0=
\{Q^\alpha,Q^\beta\}=\{\overline{Q}^\alpha,\overline{Q}^\beta\}
\qquad\textrm{with}\quad \alpha,\beta=1,\ldots,M .
\label{supalg} \ee
Here, the Hamiltonian $H_s$ differs from the previous~$H$ by nilpotent terms.
Fixing the level super-surface,
$H_s=h_s$, $Q^\alpha=q^\alpha$ and $\overline{Q}^\alpha=\overline{q}^\alpha$,
we arrive at a $(1|0)$-dimensional circle in the phase superspace.
On this circle, one defines bosonic action-angle variables
$(\Phi_s, {\tilde I}_s)$, analogous to the non-supersymmetric case,
as well as fermionic ones, $\Theta^\alpha={Q}^\alpha/\sqrt{2h_s}$,
with the following non-zero Poisson brackets~\footnote{
The tilde indicates that the action variable has been shifted as
in the previous section, depending on $k$ being odd or even.}
\be
\{\Phi_s, {\tilde I}_s\}=1 \qquad\textrm{and}\qquad
\{\Theta^\alpha,{\overline\Theta}^\beta\}=\delta^{\alpha\beta}.
\ee
In these variables, the Hamiltonian does not depend on
$\Theta^\alpha$ or ${\overline\Theta}^\alpha$, hence
$H_s={\cal I}_s({\tilde I}_s)$ just like previously.
Nevertheless, the canonical transformation from the initial to the
action-angle supervariables does mix bosonic and fermionic degrees of freedom.

Let us demonstrate the procedure for the simplest case of ${\cal N}{=}2$,
given by the classical counterpart of Witten's model of supersymmetric
mechanics~\cite{witten}. It is defined by
\be
H_s=\sfrac12\bigl(p^{2}_{\varphi}+{W'}^2(\varphi)\bigr)
+\imath\theta\overline{\theta}\,{W''}(\varphi)\qquad\textrm{and}\qquad
Q=\theta\,\bigl(p_\varphi+\imath W'(\varphi)\bigr),\quad
\overline{Q}=\overline{\theta}\,\bigl(p_\varphi-\imath W'(\varphi)\bigr),
\label{witten}\ee
with a chosen superpotential function $W(\varphi)$.
These functions obey the superalgebra~(\ref{supalg}) with~$M{=}1$, by virtue of
\be
\{p_\varphi, \varphi\}=1 \qquad\textrm{and}\qquad
\{\theta,\overline{\theta}\}=1.
\ee

Quantization replaces $\theta$ and $\overline{\theta}$ by the Pauli matrices
$\bs_+=\sfrac12(\bs_1+\imath\bs_2)$ and $\bs_-=\sfrac12(\bs_1-\imath\bs_2)$,
respectively, and $\imath\theta\overline{\theta}$ goes to $\bs_3$.
In this way we arrive at one-dimensional ${\cal N}{=}2$ supersymmetric
quantum mechanics of a spinning particle interacting with an external field.
However, when passing to action-angle variables it turns out that there is
no spin interaction, and the supersymmetric extension is rather trivial.
On the other hand, Witten's model is quite special: its supercharges
allow no momentum dependence in the nilpotent part of the Hamiltonian.
For a more interesting system related to our potentials (\ref{hggodd}) and
(\ref{hggeven}), let us choose a more flexible form of the supercharges, namely
\be
Q=\theta\,{\tilde k} {\tilde I}\,{\rm e}^{\imath\lambda({\tilde I},\Phi)}
=\sqrt{2h_s}\,\Theta
\qquad\textrm{and}\qquad
{\overline Q}=\overline{\theta}\,{\tilde k}{\tilde I}\,
{\rm e}^{-\imath\lambda({\tilde I},\Phi)}
=\sqrt{2h_s}\,\overline{\Theta},
\label{saa}\ee
where we defined
$\tilde{k}:=k$ for $k$ odd and $\tilde{k}:=k/2=k'$ for $k$ even, and
$\lambda({\tilde I},\Phi)$ is an arbitrary real function of the action-angle
variables of the underlying bosonic system. By expressing $({\tilde I},\Phi)$
through $(p_\varphi,\varphi)$, the supercharges are functions of the initial
phase superspace variables. These supercharges also generate the superalgebra
(\ref{supalg}) (with $M{=}1$) and produce the Hamiltonian
\be
H_s:=\sfrac{1}{2} \{Q,\overline{Q}\}
=\sfrac12{\tilde k}^2 {\tilde I}^2
+\imath\theta{\overline\theta}\,{\tilde k}^2 {\tilde I}\,
\frac{\partial\lambda({\tilde I},\Phi)}{\partial\Phi}.
\label{saah}\ee
The freedom of an arbitrary real function $\lambda({\tilde I},\Phi)$
leads to a variety of supersymmetric extensions of a given bosonic system.
A similar freedom
(of an arbitrary holomorphic function) has been observed in two-dimensional
${\cal N}{=}4$ supersymmetric mechanics~\cite{freedom}.

To relate to the standard ${\cal N}{=}2$ supersymmetric mechanics construction
(\ref{witten}) with $W'=\sqrt{V}$, we must choose
\be
{\tilde k}\,{\tilde I}(p_\varphi,\varphi)\,
{\rm e}^{\imath\lambda({\tilde I}(p_\varphi,\varphi),\Phi(p_\varphi,\varphi))}
=p_\varphi+\imath\sqrt{V(\varphi)},
\quad\Leftrightarrow\quad \tan\lambda=\frac{\sqrt{V(\varphi)}}{p_\varphi},
\ee
where $V(\varphi)$ is defined by (\ref{hggodd}) or~(\ref{hggeven}).
For odd $k$ we find
\be
\tan{\lambda}=\frac{\alpha_0}{\cos{\Phi}}\Bigm/\sqrt{{\tilde I}^2-\alpha_0^2},
\ee
while for even $k$ the expression is more complicated.
Note that $\lambda=\textrm{const}$ yields trivial supersymmetry,
with no spin interaction.
Another interesting case is $\lambda=\Phi/{\tilde I}$, which produces a
coordinate-independent spin-background interaction.

Applying the (super-)Liouville theorem to the supersymmetric system given by
(\ref{saa}) and (\ref{saah}), we obtain
\be
{\tilde I}_s={\tilde I}+\imath\theta{\overline\theta}\,
\frac{\partial\lambda({\tilde I},\Phi)}{\partial\Phi}, \qquad
\Phi_s=\Phi+\imath\theta{\overline\theta}\,
\frac{\partial\lambda({\tilde I},\Phi)}{\partial\tilde{I}}, \qquad
\Theta={\rm e}^{\imath\lambda({\tilde I},\Phi)}\theta,\qquad
\overline{\Theta}={\rm e}^{-\imath\lambda({\tilde I},\Phi)}\overline{\theta}.
\ee
As already said, the Hamiltonian in these variables is of the same
form as the non-supersymmetric one,
${\cal I}_s=\sfrac12{\tilde{k}}^2 {\tilde I}^2_s$.

\section{Extension to two-dimensional systems}
In any conformal mechanics one may separate the radial from the angular
degrees of freedom. The former part is universal, hence the such models
differ only by their angular Hamiltonian systems, whose coordinates
commute with the conformal algebra~$so(2,1)$~\cite{hkln}.
Such a splitting is useful for quantization~\cite{feher} and the
construction of superconformal extensions~\cite{bks,hkln}.
For $N$-particle Calogero models it yields a separation of one more variable
beyond the center of mass. Thus, their analysis becomes only complicated
starting with $N{=}4$.
For example, the angular part of the $A_{N-1}$ rational Calogero model
corresponds to a $\sfrac12N(N{-}1)$-center Higgs oscillator on $S^{N-2}$.
At $N{=}4$ its force centers are located at the vertices of a cuboctahedron
\cite{cuboct}. For $N{=}3$ however, the angular part of any rational
Calogero model lives merely on a circle, and it is precisely one of
the dihedral systems considered in this Note.

Therefore, by adding a radial coordinate $r\in[0,\infty)$,
we may extend our one-dimensional system to a two-dimensional
conformal mechanics with dihedral symmetry
(a rational 3-particle Calogero model),
defined by the $SO(2,1)$ generators
\be
{\cal H}_0=
\frac{{p}^{2}_r}{2}+ \frac{{\cal I}(\tilde{I})}{r^2}=
\frac{{\bf p}^2}{2}+ \sum_{\ell=0}^{k-1}\frac{1}{({\bf a}_\ell\cdot{\bf r})^2},
\qquad {\cal D}=p_r r={\bf pr} ,
\qquad {\cal K}=\sfrac12 r^2=\sfrac12{\bf r}^2,
\label{h20}\ee
where ${\bf a}_\ell$ run over the positive dihedral roots as before.
This allows us to extend the above-established equivalence of systems with
different $k$-values to these two-dimensional systems.
In particular, all these Calogero models are {\it locally\/} equivalent
to a free particle on the plane,
which is in agreement with the ``decoupling'' transformation of
the quantum Calogero model to the free particle~\cite{decoupling}.
Furthermore, since for certain small values of~$k$ the model is based
on a Lie algebra listed in~(\ref{tab}),
we can also assert the {\it global\/} equivalence of the $G_2$~model with
couplings ($\alpha_1,\alpha_2$) to the~$A_2$ model with coupling
$\alpha_0=\alpha_1+\alpha_2$.

Since the radial motion is unbounded,
the Hamiltonian (\ref{h20}) does non admit a formulation in terms of
action-angle variables. This complication may be avoided by adding
an oscillator potential,
\be
{\cal H}=\frac{{\bf p}^2}{2}+
\sum_{\ell=0}^{k-1}\frac{1}{({\bf a}_\ell\cdot {\bf r})^2} +
\frac{\omega^2{\bf r}^2}{2}=
\frac{{p}^{2}_r}{2}+ \frac{{\cal I}(\tilde{I})}{r^2}+\frac{\omega^2r^2}{2}.
\label{h2}\ee
The confining potential allows for the application of the Liouville theorem.
Thus, in order to extend the action-angle variable formulation
to the latter system, we fix the level surface of the constants of motion 
${\cal H}$ and $\tilde{I}$ and introduce the generating function
in accordance with the expression for the symplectic one-form 
${\tilde I}d\Phi+p_rdr$,
\be
S\ =\ {\tilde I}\,\Phi+\int_{r_{0}}^{r}\!dr'\
\sqrt{2h-\sfrac{2{\cal I}({\tilde I})}{{r'}^2} -{\omega^2{r'}^2}}
\ =\ {\tilde I}\,\Phi+\int_{r_{0}}^{r}\!dr'\
\sqrt{2h-\sfrac{(\tilde{k}{\tilde I})^2}{{r'}^2} -{\omega^2{r'}^2}},
\ee
where $h$ is the value of the Hamiltonian~${\cal H}$.
By the standard technique, we identify the action variables as
\be
I_{\textrm{ang}}={\tilde I} \qquad\textrm{and}\qquad
I_{\textrm{rad}}=\frac{h}{2\omega}-\frac{\tilde k\tilde I}{2}
\label{av}\ee
and find the canonically conjugated angle variables
\be
\Phi_{\textrm{ang}}=
\Phi+{\tilde k}\Phi_{\textrm{rad}} -
\arcsin\sfrac{h-\frac{{\tilde k}{\tilde I}}{r^2}}{\sqrt{h^2-{\tilde k}^2}}
\qquad\textrm{and}\qquad
\Phi_{\textrm{rad}}=
-\arcsin\sfrac{h-\omega^2r^2}{\sqrt{h^2-({\tilde k}{\tilde I}\omega)^2}}.
\ee
Equation~(\ref{av}) gives us the Hamiltonian in terms of action variables,
\be
{\cal I}=\omega\,(2{I}_{\textrm{rad}}+{\tilde k}{I}_{\textrm{ang}}).
\ee
One sees that our confined system~(\ref{h2}) is locally equivalent to a
two-dimensional anisotropic oscillator with frequencies 
$\omega_{\textrm{rad}}=\omega$ and
$\omega_{\textrm{ang}}=\sfrac12{\widetilde k}\omega$. 
Since the frequency ratio is
rational, the trajectories on the two-torus are closed.

Finally, we note that the general approach for the construction of
action-angle variables of Calogero models has been presented in~\cite{ruij}.
There, the action variables are associated with the Lax constants of motion.
On the other hand, due to its superintegrability, the Calogero model enjoys
an additional series of constants of motion~\cite{supint}.
Our construction of action-angle variables is in fact related to these
additional constants of motion.
In the $A_2$ Calogero model, for instance, the angular Hamiltonian ${\cal I}$
is a function of the Lax constants of motion as well as the Woijechowski one
\cite{cuboct}.

\subsection*{Conclusion}
In this Note we demonstrated, on the example of the one-dimensional dihedral 
systems, that action-angle variables are an effective tool
for establishing the (non)equivalence of any two integrable systems.

For this purpose we gave the action-angle formulation of these one-particle 
systems on the circle, interacting via the Higgs oscillator law~\cite{higgs}, 
with $2k$ equally spaced force centers at the Coxeter root systems~$I_2(k)$.
We established the local equivalence of any of their $2k$ branches with a 
free particle on the circle,
as well as a global equivalence of any two systems with 
$k_{\textrm{even}}=2k_{\textrm{odd}}$
upon an appropriate identification of their couplings.

Besides one particle on a circle, the dihedral systems represent also the
angular part of the three-particle rational Calogero models (after stripping 
off the center of mass), based on the same root systems.
This fact allowed us to prove the global equivalence of the $A_2$ and $G_2$ 
rational Calogero model as the simplest example, 
and qualifies the known equivalence of the Calogero model to a free-particle
system as a local one.
Advancing to higher-rank models should yield a separation of variables for 
rational Calogero models with more than three particles.
This is highly non-trivial, since technically it amounts to solving
fourth-order algebraic equations.

The action-angle formulation is also helpful for the construction of
supersymmetric extensions: due to the absence of ``interaction terms", 
the supersymmetrization itself is rather trivial. The non-triviality lies
in the construction of the supercanonical transformation from the action-angle
variables to the initial phase (super)space variables and vice versa.
We demonstrated how to define such a supercanonical transformation for the
example of ${\cal N}{=}2$ supersymmetric mechanics. 
Unexpectedly, there exists a functional freedom in the supersymmetrization 
procedure, which yields a family of non-equivalent supersymmetric extensions.
This feature and its possible analogs in ${\cal N}{=}4$ and ${\cal N}{=}8$
supersymmetric extensions of action-angle variables merit further investigation.

\bigskip
\noindent
{\large Acknowledgments.}\\
We are  grateful to Tigran Hakobyan  for useful discussions and comments.
The work was supported by and ANSEF-2229PS grant and by
Volkswagen Foundation  grant I/84~496.

\subsection*{Appendix: Calculation of the integral (\ref{idef2})}
To compute the value of the definite integral in (\ref{idef2})
we perform the substitution
\be
\sqrt{1-x^2}=x\,t-1 \qquad\Leftrightarrow\qquad
x=\frac{2t}{1+t^2}.
\ee
In this variable the integral reads
\be
\int\limits_{-1}^1 \frac{-2(1-t^2)\ dt}
{(1+t^2)\left((1-b)t^2-2at+1-b\right)\left((1+b)t^2+2at+1+b\right)}.
\ee
Decomposing the integrand into partial fractions we get
\be
\int\limits_{-1}^1 \frac{dt}{a^2 \left(1+t^2\right)}-\frac{1}{2 a^2(1-b)}
\int\limits_{-1}^1\frac{a (1- b) t+(1-b)^2-2 a^2}{ t^2-\frac{2 a}{(1-b)} t+1}dt
+\frac{1}{2 a^2(1+b)}\int\limits_{-1}^1\frac{a (1+ b) t-(1+b)^2+2 a^2}
{ t^2+\frac{2 a}{(1+b)} t+1}dt.
\ee
The first integral yields $\frac{\pi}{2}$, while
for the second and the third term we obtain respectively
\be
-\left.\frac{1}{2 a^2}\sqrt{((1-b)^2- a^2)}\arctan{\frac{(1-b)t-1}{\sqrt{(1-b)^2-a^2}}}\right|_{-1}^1
\qquad\textrm{and}\qquad
-\left.\frac{1}{2 a^2}\sqrt{(1+b)^2-a^2})\arctan{\frac{(1+b)t +a}{\sqrt{(1+b)^2-a^2}}}\right|_{-1}^{1}.
\ee
Finally, using the elementary relation
\be
\arctan{x}-\arctan{y}=\arctan{\frac{x-y}{1+xy}},
\ee
and combining all terms we arrive at
\be
\int\limits_{-1}^1\frac{dx\ \sqrt{1-x^2}}{1-\left(ax+b\right)^2}
=\frac{\pi}{4a^2}\left(2-\sqrt{(1-b)^2- a^2}-\sqrt{(1+b)^2-a^2}\right) .
\ee

\end{document}